\pdfoutput=1
\documentclass[aps,prl,10pt,twocolumn,superscriptaddress]{revtex4}
\usepackage{mathrsfs, amssymb, amsmath, amsfonts, txfonts, latexsym, graphicx}
\usepackage[T1]{fontenc}
\usepackage[
      colorlinks=true,
      linkcolor=blue,
      urlcolor=cyan,
      filecolor=magenta,
      citecolor=red,
      pdfstartview=FitV,
      hyperfootnotes=false,
      pdftitle={Dynamics of Carrollian Scalar Fields},
        pdfauthor={Luca Ciambelli},
        pdfsubject={Dynamics of Carrollian Scalar Fields},
        pdfkeywords={Dynamics of Carrollian Scalar Fields},
        bookmarksopen=true
      ]{hyperref}
\usepackage{cancel}
\usepackage{mdframed,framed}
\usepackage[usenames,dvipsnames]{xcolor}
\usepackage{layout}

\newcommand{\be}{\beta}

\newcommand{\beq}{\begin{eqnarray}}
\newcommand{\eeq}{\end{eqnarray}}
\newcommand{\beqn}{\begin{eqnarray}}
\newcommand{\eeqn}{\end{eqnarray}}
\newcommand{\pa}{\partial}

\newcommand{\cL}{{\cal L}}
\newcommand{\ve}{{\varepsilon}}
\newcommand{\cC}{{\cal C}}

\def\be#1\ee{\begin{align}#1\end{align}}
\newcommand{\ee}{\end{equation}}
\newcommand{\bea}{\begin{eqnarray}}
\newcommand{\eea}{\end{eqnarray}}
\newcommand{\bs}{\begin{subequations}}
\newcommand{\es}{\end{subequations}}
\newcommand{\nn}{\nonumber}

\newcommand{\cM}{{\cal M}}

\DeclareFontFamily{OT1}{rsfs}{}
\DeclareFontShape{OT1}{rsfs}{m}{n}{ <-7> rsfs5 <7-10> rsfs7 <10->rsfs10}{} 
\DeclareMathAlphabet{\mycal}{OT1}{rsfs}{m}{n}

\begin{document}


\title{Dynamics of Carrollian Scalar Fields}

\author{Luca Ciambelli}
\affiliation{Perimeter Institute for Theoretical Physics, 31 Caroline St. N., Waterloo ON, Canada, N2L 2Y5\\
\href{mailto:ciambelli.luca@gmail.com}{ciambelli.luca@gmail.com}}

\begin{abstract}
Adopting an intrinsic Carrollian viewpoint, we show that the generic Carrollian scalar field action is a combination of electric and magnetic actions, found in the literature by taking the Carrollian limit of the relativistic scalar field. This leads to non-trivial dynamics: even a single particle with non-vanishing energy can move in Carrollian physics. 
\end{abstract}

\maketitle


\section{Introduction}

Since its discovery by L\'evy-Leblond \cite{Levy1965} and Sen Gupta \cite{Gupta1966}, the Carrollian universe -- found taking the $c\to 0$ contraction of relativistic physics -- keeps surprising us with unexpected and unusual features. By now, there have been numerous avenues of investigation in this field, such as its relationship to the BMS group \cite{Duval:2014lpa, Duval:2014uva, Duval:2014uoa,  Ciambelli:2019lap}, the fluid/gravity correspondence and hydrodynamics \cite{deBoer:2017ing, Ciambelli:2018xat, Ciambelli:2018ojf, Campoleoni:2018ltl, Ciambelli:2020eba, Ciambelli:2020ftk, Freidel:2022bai, Petkou:2022bmz}, flat holography and Carrollian field theory \cite{Bagchi:2016bcd, Ciambelli:2018wre, Bagchi:2019xfx, Bagchi:2019clu, Gupta:2020dtl, Banerjee:2020qjj, Figueroa-OFarrill:2021sxz, Chen:2021xkw, Herfray:2021qmp, Donnay:2022aba, Bagchi:2022eav, Bekaert:2022oeh, Bagchi:2022emh, Rivera-Betancour:2022lkc, Baiguera:2022lsw, Campoleoni:2022wmf, Donnay:2022wvx, Dutta:2022vkg, Mittal:2022ywl,Chen:2023pqf, Mehra:2023rmm, Campoleoni:2023fug, Salzer:2023jqv, Bagchi:2023fbj, Saha:2023hsl, Banerjee:2023jpi}, black hole horizon and generic null hypersurfaces \cite{Penna:2015gza, Penna:2018gfx, Donnay:2019jiz, Redondo-Yuste:2022czg, Freidel:2022vjq, Gray:2022svz, Ciambelli:2023mir, Ciambelli:2023mvj}, Carrollian particles and algebras \cite{Bergshoeff:2014jla, Hartong:2015xda, Henneaux:2021yzg, Marsot:2022imf, Bergshoeff:2022eog, Bagchi:2022eui, Kasikci:2023tvs, Casalbuoni:2023bbh, Cerdeira:2023ztm, Zhang:2023jbi, Kamenshchik:2023kxi}, Carrollian gravity \cite{Henneaux:1979vn, Bergshoeff:2017btm, Matulich:2019cdo, Gomis:2020wxp, Grumiller:2020elf, Concha:2021jnn, Perez:2021abf, deBoer:2021jej, Hansen:2021fxi, Campoleoni:2022ebj, Ekiz:2022wbi, Figueroa-OFarrill:2022mcy, Miskovic:2023zfz} and black holes \cite{Ecker:2023uwm}, supersymmetry and supergravity \cite{Ravera:2019ize, Ali:2019jjp, Ravera:2022buz, Kasikci:2023zdn}, Carrollian strings \cite{Schild:1976vq, Isberg:1993av, Bagchi:2015nca, Fursaev:2023lxq, Fursaev:2023oep}, and fractons \cite{Bidussi:2021nmp, Figueroa-OFarrill:2023vbj, Figueroa-OFarrill:2023qty, Perez:2023uwt}. See \cite{deBoer:2023fnj} and references therein for a recent account on Carrollian physics.

Here, rather than obtaining Carrollian physics as usually done via a limiting procedure, we adopt an intrinsic perspective and construct a Carrollian physical theory compatible with symmetries and geometry. It is reasonable to expect that the intrinsic approach reveals physical scenarios that are not reachable through a limiting procedure from another theory. Indeed, we show in this paper that this is the case for the simplest model, the Carrollian scalar field. This has been vastly studied in the literature, see \cite{Ciambelli:2018ojf, Bagchi:2019xfx, Gupta:2020dtl, Chen:2021xkw, Bekaert:2022oeh, Bagchi:2022eav, Chen:2023pqf, Banerjee:2023jpi, Casalbuoni:2023bbh, Zhang:2023jbi, deBoer:2023fnj}. In particular, it has been recently discussed in the papers of Rivera-Betancour and Vilatte (RBV) \cite{Rivera-Betancour:2022lkc}, and Baiguera, Oling, Sybesma, and Søgaard (BOSS) \cite{Baiguera:2022lsw}. In both works, the (conformal) Carrollian scalar field has been obtained as the limit of the relativistic conformal scalar field. It has also been shown that the so-called electric and magnetic actions cannot coexhist, albeit for different reasons. In RBV, these two actions appear at different powers in $c\to 0$, and so must be considered separately. In BOSS, the magnetic action is required to be local Carroll boost invariant, which is the reason why their timelike (electric) and spacelike (magnetic) actions do not mix. 

This work is based on the simple observation -- with dramatic consequences -- that these electric and magnetic actions coexist intrinsically on a Carrollian manifold, without invoking a limiting procedure. This has been already contemplated using Carrollian diffeomorphisms in \cite{Gupta:2020dtl}, and in two dimensions in \cite{Bagchi:2022eav}. Here, we will study so retaining the full diffeomorphism invariance. Indeed, the magnetic action we consider is invariant under diffeomorphisms, but not under internal frame Carroll boost as considered by BOSS. We accept this in order to uncover non-trivial dynamics, and postpone its consequences to future studies. In a companion paper, Grumiller and the author \cite{Ciambelli:2023tzb} explored a seemingly disconnected yet ultimately related problem, the study of geodesics in Carrollian gravity (intrinsically). We show that a strikingly similar result holds there: the temporal and spatial parts of a particle trajectory mix together and lead to geodesic motion. Therefore, the presence of dynamics in Carrollian physics seems to be a common feature. 

Our result indicates that $c^2$ might be treated as an intrinsic Carrollian coupling constant for the scalar field, while it ceases to be interpreted as the speed of light for the background. We show that the equation of motion is nothing but the Klein-Gordon equation, with only one difference: the speed of light is replaced by the relative coupling constant between the two actions. This means that a Carrollian scalar field undergoes non-trivial dynamics, but the dispersion relation is determined by the way the electric and magnetic parts are  coupled together. In the simplest geometric setup, we construct the Feynman's Green function and show that there are wave solutions with non-trivial dispersion relations, such that the Carrollian vacuum can be understood as a non-trivial medium. We defer to further studies the situation on a general background, and the implications for flat holography. The former inevitably involves technicalities concerning Carrollian connections (that we overview in Appendix \ref{AppA}), while the latter, which is one of our main motivations, would require us to revisit the extrapolating dictionary. We want here to emphasize our main result alone: while motion for zero energy particles has been observed in \cite{deBoer:2023fnj}, and multi-particles motion in \cite{Casalbuoni:2023bbh, Zhang:2023jbi}, this is the first time that non-trivial motion is observed for a single free Carrollian scalar field, with arbitrary energy.

\section{Geometric Setup}\label{s2}

Consider a $d+1$-dimensional manifold $\cM$. 
A Carrollian structure is determined by a nowhere vanishing Carrollian vector $v=v^\mu\pa_\mu$ and a rank-$d$ (co-rank $1$) degenerate metric $h_{\mu\nu}$ satisfying $v^\mu h_{\mu\nu}=0$. A ruled Carrollian structure has an extra ingredient, the Ehresmann connection, which is a one-form $E=E_\mu\text{d}x^\mu$ satisfying $E_\mu v^\mu=1$. This object is defined modulo shifts  in the kernel of $v^\mu$
\beq\label{boo}
E'_\mu=E_\mu-\lambda_\mu \quad \text{with} \quad \lambda_\mu v^\mu=0.
\eeq
Such a shift is called a Carroll boost. It has two possible origins. 

The first one is as an internal frame-bundle transformation, as done in \cite{Hartong:2015xda, Hansen:2021fxi, Baiguera:2022lsw}. In this case, this transformation changes the field but does not act on the coordinates. The vector $v^\mu$ and the degenerate metric $h_{\mu\nu}$ are invariant under this internal boost.
We will call this transformation a {\it  local Carroll boost}. 

A different origin of the Carroll boost is as a diffeomorphism, a covariantization of the usual coordinate boost ($x^\mu=(t,x^i)$)
\beq
t'=t+\lambda_i x^i\qquad x'^i=x^i,
\eeq
which reads
\beq
x'^\mu\simeq x^\mu+x^\nu\lambda_\nu v^\mu.
\eeq
This is a diffeomorphism generated by $\xi^\mu=-v^\mu \lambda_\alpha x^\alpha$, and thus
\beqn
E'_\mu- E_\mu\simeq \cL_{\xi}E_\mu=-\lambda_\mu-2v^\alpha\lambda_\gamma x^\gamma F_{\alpha\mu}-x^\alpha\pa_\mu \lambda_\alpha,
\eeqn
where we introduced the Carrollian field strength \footnote{Related to the usual Carrollian vorticity by
\beq
F_{\alpha\mu}=\omega_{\alpha\mu}-\frac12(E_\alpha a_\mu-E_\mu a_\alpha),
\eeq
where $a_\mu=v^\nu D_\nu E_\mu$ is the Carrollian acceleration, see \cite{Ciambelli:2018wre}.}
\beq
F_{\alpha\mu}=\frac12(\pa_\alpha E_\mu-\pa_\mu E_\alpha).
\eeq
So we recover eq. \eqref{boo} under this change of coordinates assuming $2v^\alpha\lambda_\gamma x^\gamma F_{\alpha\mu}=-x^\alpha\pa_\mu \lambda_\alpha$. We refer to this change of coordinates as a {\it diff Carroll boost}. The action on all the fields is the same for both diff and local Carroll boosts if we further require  $v^\mu\pa_\mu\lambda_\alpha=0$ and $\cL_{\xi}h_{\mu\nu}=x^\alpha \lambda_\alpha \cL_{v}h_{\mu\nu}=0$. While these conditions are needed to reproduce \eqref{boo}, {\it we will not impose it in the following}, as our construction will be fully diffeomorphism invariant. 

The Ehresmann connection allows us to specify the (boost-invariant) volume form $\ve_{\cM}=E \wedge \ve_{\cC}$, where $\cC$ is a cut of $\cM$ transverse to $v^\mu$.
The degenerate metric, Ehresmann connection, and Carrollian vector field allow us to introduce the projector
\beq
h_\mu{}^\nu=\delta_\mu^\nu-E_\mu v^\nu \qquad v^\mu h_\mu{}^\nu=0=h_\mu{}^\nu E_\nu,
\eeq
which can be used to raise and lower indices in the space orthogonal to $v^\mu$. We can use this to introduce the symmetric tensor $h^{\mu\nu}$, defined via
\beq\label{c1}
E_\mu h^{\mu\nu}=0 \qquad h_{\mu\rho}h^{\rho\nu}=h_\mu{}^\nu.
\eeq
Both the projector and the tensor $h^{\mu\nu}$ transform non-trivially under (local and diff) Carroll boosts. Under the conditions spelled above, we have
\beqn
\cL_\xi h_\mu{}^\nu&= &h_\mu{}^\nu+\lambda_\mu v^\nu\\
\cL_\xi h^{\mu\nu}&= &h^{\mu\nu}+v^\mu\lambda^\nu+v^\nu\lambda^\mu,\label{bq}
\eeqn
where we defined $\lambda^\mu=h^{\mu\nu}\lambda_\nu$. 

There is no analogue of the Levi-Civita connection for a ruled Carrollian structure. We relegate a full account of Carrollian connections to Appendix \ref{AppA}, and we here require the degenerate metric $h_{\mu\nu}$ to be time-independent $v(h_{\mu\nu})=v^\alpha\pa_\alpha h_{\mu\nu}=0$, for which  torsion-free and metricity can be imposed. This choice is motivated by three reasons. First of all, we will later impose the simplest possible setup, as it will be enough to unravel the leitmotif of this paper, that is, that the Carrollian scalar field is dynamical. Secondly, we prove in \cite{Ciambelli:2023tzb} that Carrollian geodesics also have non-trivial dynamics, and we test that on simple geometric backgrounds, where this assumption holds. Eventually, this assumption also persists in celestial holography applications, where the Carrollian scalar field lies on null infinity ${\cal I}$, and Einstein equations impose geometric constraints on ${\cal I}$. Therefore, we assume that our Carrollian connection satisfies the torsion-free condition, together with
\beq\label{conn1}
&D_\mu h_{\nu\rho}=0 \quad (D_\mu-\omega_\mu) v^\nu=0& \\
& (D_\mu+\omega_\mu) \ve_{\cM}=0\quad (D_\mu+\omega_\mu) E_\nu=0,&\label{conn2}
\eeq
where we introduced
\beq
\omega_\mu=\kappa E_\mu+\pi_\mu,\qquad \pi_\mu v^\mu=0.
\eeq
See \cite{Chandrasekaran:2021hxc, Freidel:2022vjq, Ciambelli:2023mir} for more details \footnote{Dictionary for the comparison with \cite{Ciambelli:2023mir}: $x^\mu \leftrightarrow x^a$, $v^\mu \leftrightarrow \ell^a$, $h_{\mu\nu} \leftrightarrow q_{ab}$, and $E_\mu \leftrightarrow k_a$.}. Note that we are assuming that $v^\mu$ is a Killing vector, $\cL_v h_{\mu\nu}=0$.

Finally, our connection parallel transports the projector and $h^{\mu\nu}$,
\beq
D_\mu h_\nu{}^\rho=0\qquad D_\mu h^{\nu\rho}=0.
\eeq

\section{Scalar Field}

We wish to construct a free Carrollian scalar field action on a general background, where we require full diffeomorphism invariance and only assume from now on $v^\alpha\pa_\alpha h_{\mu\nu}=0$. We focus here on the massless scalar field.
Given the objects at our disposal, there are $2$ independent terms that can be written for a  scalar field action containing $2$ derivatives of the field \footnote{Notice that a priori, being the action a sum of two terms, there could be two independent scalar fields appearing. It is trivial to see, however, that one can add a Lagrange multiplier that makes these two fields proportional.}
\beq\label{ac}
S[\phi]=-\int_{\cM}\ve_{\cM} \left(g_1 v^\mu\pa_\mu\phi v^\nu\pa_\nu \phi+ g_2 h^{\mu\nu}\pa_\mu \phi \pa_\nu\phi\right),
\eeq
where $g_1$ and $g_2$ are independent coupling constants. These could be functionals of the field, but in the rest we assume that they are not.
Assuming $g_2\neq 0$ and introducing $g_0=-\frac{g_1}{g_2}$, we get 
\beqn
S[\phi]&=&g_2\int_{\cM}\ve_{\cM} \left(g_0 v^\mu\pa_\mu\phi v^\nu\pa_\nu \phi- h^{\mu\nu}\pa_\mu \phi \pa_\nu\phi\right). \
\eeqn
The first term is the kinetic term in the timelike action derived by BOSS, while the second term is nothing but their free (i.e., without constraints) spacelike action. Notice that BOSS and RBV studied conformal Carroll scalar fields, so they have extra terms in the action to ensure conformality. While classically one can assume that $g_0$ has definite sign -- which we will assume positive --, this is a dimensionful coupling constant, and thus it can run in the quantum theory. One should then be careful that it does so keeping definite sign, in order for our discussion later to be well-defined.

The electric action
\beq
S_e[\phi]=-g_1\int_{\cM}\ve_{\cM}  v^\mu\pa_\mu\phi v^\nu\pa_\nu \phi
\eeq
is manifestly local Carroll boost invariant and diffeomorphism invariant.
The magnetic action 
\beq
S_m[\phi]=-g_2\int_{\cM}\ve_{\cM}h^{\mu\nu}\pa_\mu \phi \pa_\nu\phi
\eeq
is not local Carroll boost invariant, as proven by BOSS, but it is by construction diffeomorphism invariant. Aware of the non-invariance of the magnetic action under local Carroll boosts, we continue our analysis, as our result will be non-trivial and is supported by the analysis in \cite{Ciambelli:2023tzb}. A full account of the symmetries has been done by BOSS, and pertains the electric and magnetic actions separately. For this reason, we do not report details here, and refer to \cite{Baiguera:2022lsw} for more information.

We then remark that the actions  studied by \cite{Gupta:2020dtl} and  RBV are also included in this analysis. To see this, one introduces coordinates $x^\mu=(t,x^i)$, and uses the Papapetrou-Randers parameterization
\beq\label{rp}
v^\mu\pa_\mu=\frac1{\Omega}\pa_t \quad h_{\mu\nu}=a_{ij}\delta^i_\mu\delta^j_\nu\quad E_\mu\text{d}x^\mu=\Omega \text{d}t-b_i\text{d}x^i.
\eeq
From this, defining $a^{ij}$ as the inverse of the (non-degenerate) tensor $a_{ij}$, and $b^i=a^{ij}b_j$, we derive
\beq\label{rpi}
h^{\mu\nu}=\begin{pmatrix}
    \frac{b^k b_k}{\Omega^2} & \frac{b^j}{\Omega}\\
    \frac{b^{i}}{\Omega} & a^{ij}
\end{pmatrix}.
\eeq
Notice that our assumption that $v^\mu$ is a Killing vector becomes in this parameterization $\pa_t a_{ij}=0$, but this assumption is not required in the comparison with RBV.
It is then a straightforward exercise to see that our action \eqref{ac} reduces to the electric and magnetic actions of RBV. As shown in \cite{deBoer:2021jej} and by RBV, these two actions appear at different $c$-orders in the Carrollian limit. Indeed, consider the action for the free relativistic scalar field theory on a non-degenerate flat background
\beq\label{rel}
S[\phi]=\int c\text{d}t \text{d}^d x \left(\frac1{2c^2}(\pa_t\phi)^2-\frac12 \delta^{ij}\pa_i\phi \pa_j\phi\right).
\eeq
The limit $c\to 0$ yields the electric and magnetic actions, at different powers of $c$. Therefore, as already stated, these two actions cannot coexist in the Carrollian limit.
The main message of this paper is that they can, nevertheless, arise together from an intrinsic Carrollian viewpoint. 

With our time-independent condition on $h_{\mu\nu}$, the equations of motion can be derived on an otherwise general background. Linearity of the action allows us to focus on the two sectors separately. We start with the electric sector. Using \eqref{conn1} and \eqref{conn2}, we get ($\pa_\mu\phi=D_\mu\phi$)
\beqn
\delta S_e[\phi]&=&-2g_1\int_{\cM}\ve_{\cM}v^\mu D_\mu(\delta\phi) v^\nu D_\nu\phi\nn\\
&=&-2g_1\int_{\cC}\ve_{\cC}\delta\phi v^\mu D_\mu \phi+2g_1\int_{\cM}\delta\phi D_\mu (\ve_{\cM}v^\mu v^\nu D_\nu\phi)\nn\\
&=&-2g_1\int_{\cC}\ve_{\cC}\delta\phi v^\mu D_\mu \phi\nn\\
&&+2g_1\int_{\cM}\ve_{\cM}(v^\mu v^\nu D_\mu D_\nu\phi+\kappa v^\mu D_\mu\phi),
\eeqn
where we used $v^\mu\omega_\mu=\kappa$, and chose the cut $\cC$ orthogonal to $v^\mu$. From $\delta S[\phi]=\Theta+\int_{\cM}\ve_{\cM}EOM \delta \phi$, we read
\beq\label{te}
&\Theta_e=-2g_1\int_{\cC}\ve_{\cC}\delta\phi v^\mu D_\mu \phi&\\
&EOM_e=2g_1(v^\mu v^\nu D_\mu D_\nu\phi+\kappa v^\mu D_\mu\phi).&
\eeq
Note that the equations of motion can be rewritten in another suggestive way
\beq
EOM_e=2g_1 v^\mu D_\mu(v^\nu D_\nu[\phi]).
\eeq

For the magnetic sector, we have
\beq
\delta S_m[\phi]&=&-2g_2\int_{\cM}\ve_{\cM}h^{\mu\nu}D_\mu(\delta\phi)D_\nu\phi\\
&=&2g_2\int_{\cM}\ve_{\cM} \delta\phi(h^{\mu\nu}D_\mu D_\nu\phi-h^{\mu\nu}\pi_\mu D_\nu \phi),\nn
\eeq
where we used that the cut is orthogonal to $v^\mu$, and thus there are no contributions to the symplectic potential \footnote{Had we not assumed that, we would have had a contribution to the symplectic potential given by
\beq
-2g_2\int_{\cC} \ve_\mu \delta \phi h^{\mu\nu} D_\nu\phi
\eeq
where $\ve_\mu$ is the induced volume to the cut. Our choice of cut is such that $v^\mu \ve_\mu=\ve_{\cC}$ and $h^{\mu\nu}\ve_\mu=0$.}. Therefore we have found
\beq
\Theta_m=0 \quad EOM_m=2g_2(h^{\mu\nu}D_\mu D_\nu\phi-h^{\mu\nu}\pi_\mu D_\nu \phi).
\eeq

Putting the two contributions together, and using $g_0=-\frac{g_1}{g_2}$, we get the equations of motion of our Carrollian scalar field
\beq\label{eom}
(g_0 v^\mu v^\nu-h^{\mu\nu})D_\mu D_\nu\phi+(g_0\kappa v^\mu+\pi_\nu h^{\mu\nu})D_\mu\phi=0,
\eeq
on a background that has time-independent degenerate metric $h_{\mu\nu}$. This equation clearly indicates that there is non-trivial motion in this theory. The first parenthesis is reminiscent of the relativistic Klein-Gordon operator, except that the speed of light is replaced by (the inverse of the square root of) $g_0$. It is as if there is a fictitious non-degenerate inverse metric 
\beq
g^{\mu\nu}=-g_0 v^\mu v^\nu+h^{\mu\nu}.
\eeq
The second parenthesis in \eqref{eom} is dictated by the non-trivial connection $D_\mu$: although $h_{\mu\nu}$ is parallel transported by $D_\mu$, $v^\mu$ is not, and these terms keep that into account. They are useful to understand the propagation of Carrollian fields on a curved background. Although we set them to zero in the Section, we plan to focus on them in future works, for the derivation of the equations of motion on a completely general background is currently under investigation.

We conclude the Section with a comment on the symplectic structure. With our choice of transverse cut $\cC$, the symplectic potential is \eqref{te}. The symplectic $2$-form $\Omega=\delta\Theta$ is thus
\beq
\Omega=2g_0 g_2\int_{\cC}\ve_{\cC}v^\mu D_\mu \delta\phi\wedge\delta\phi.
\eeq
This means that the symplectic data are $\phi$ and its "temporal" derivative $v^\mu\pa_\mu\phi=\dot\phi$, and the equal-time kinematic Poisson bracket is
\beq
\{\dot\phi(x_1),\phi(x_2)\}=\frac1{2g_0g_2}\delta^{(d)}(x_1-x_2),
\eeq
where the covariant $\delta$ function is defined via
\beq
\int_{\cC} \ve_{\cC} \delta^{(d)}\left(x_1-x_2\right) f\left(x_1\right)=f\left(x_2\right).
\eeq

\section{Simplest Setup}

While the dynamics is clear already from \eqref{eom}, we want to make further simplifying assumptions, in order to reach its principal part. We impose that the background is endowed with a strong Carroll structure, as originally defined in \cite{Duval:2014lpa, Duval:2014uva, Duval:2014uoa}. This implies that both $h_{\mu\nu}$ and $v^\mu$ are parallel transported, and thus it further imposes
\beq
\kappa=0\qquad \pi_\mu=0.
\eeq

We then choose the vector field and degenerate metric to be flat, that is, using coordinates $x^\mu=(u,x^i)$ \footnote{The null time coordinate is typically called $u$ on null infinity ${\cal I}$ and $v$ on a black hole horizon. With future applications to celestial holography in mind, we use $u$ here.},
\beq\label{fl}
v^\mu\pa_\mu=\pa_u\qquad h_{\mu\nu}=\delta_{ij}\delta^i_\mu\delta^j_\nu.
\eeq
Note that on this background we can use the diff Carroll boost invariance to fix the Ehresmann connection to $E=\text{d}u$. This implies
\beq
h^{\mu\nu}=\delta^{ij}\delta_i^\mu\delta_j^\nu,
\eeq
but, while \eqref{fl} is, this statement is not (local and diff) Carroll boost invariant. Under the latter $h^{\mu\nu}$ transforms as in \eqref{bq} and thus $\delta_{ij}$ is boost invariant but $\delta^{ij}$ is not, as part of the degenerate tensor $h^{\mu\nu}$ \footnote{A simple way to see this is to perform a generic Carrollian diffeomorphism $u'(u,x)$ and $x'(x)$, as defined in \cite{Ciambelli:2018xat, Ciambelli:2018wre}. Calling the diffeomorphism generators $\Omega(u,x)$ and $b_i(u,x)$, one obtains \eqref{rp} (with $t\leftrightarrow u$), and in particular \eqref{rpi}, which shows the transformation rule of $\delta^{ij}$. Note that the finite form of \eqref{bq} is 
\beq
h'^{\mu\nu}=h^{\mu\nu}+v^\mu\lambda^\nu+v^\nu\lambda^\mu+\lambda_\alpha \lambda^\alpha v^\mu v^\nu.
\eeq}. 

Under all these assumptions, the Carroll scalar action becomes
\beq\label{cs}
S=g_2\int_{\cM}\text{d}u\text{d}^d x \left(g_0 \pa_u\phi \pa_u \phi- \delta^{ij}\pa_i \phi \pa_j\phi\right).
\eeq
Its equations of motion are trivial to derive, either by direct computation or using \eqref{eom} \footnote{This action can be derived from the construction in \cite{Gupta:2020dtl}, taking simplifying assumptions. Their starting point is a Carroll-diffeomorpshims invariant action, while our is a full diffeormorphisms invariant one.}
\beq
g_0\pa_u^2\phi-\delta^{ij}\pa_i\pa_j\phi=0.
\eeq
We finally reached the main result of this manuscript. As advertised, the equation of motion is simply the Klein-Gordon equation, where the speed of light is replaced by the relative coupling constant between the electric and magnetic sectors $c\leftrightarrow \frac1{\sqrt{g_0}}$. This is directly seen comparing \eqref{cs} with \eqref{rel}. While in relativistic physics this coupling constant is the speed of light of the background, here the latter is not defined, but, remarkably, this does not prevent motion. Notice also that $g_0$ could be negative or positive. The negative branch could be understood as an effective Euclidean scalar, where instead of Wick-rotating time one makes the speed of light imaginary. Here we are mostly interested in the positive branch.

Calling $x^\mu=(u,x^i)$ and $x'^\mu=(u',x'^i)$, the Green's function for this scalar field  satisfies
\beq
i(g_0\pa_u^2-\delta^{ij}\pa_i\pa_j)G(x-x')=\delta^{(d+1)}(x-x').
\eeq
Its Fourier transform is given by
\beq
G(x-x')=\int \frac{\text{d}^{d+1} k}{(2\pi)^{d+1}}e^{-i k_\mu (x^\mu-x'^\mu)}\tilde{G}(k),
\eeq
where $k_\mu=(k_u,k_i)$ is the $d+1$ Carrollian momentum. Therefore, we have \footnote{Conventions \beq
\delta^{(d+1)}(x-x')=\int \frac{\text{d}^{d+1}k}{(2\pi)^{d+1}}e^{ik_\mu (x^\mu-x'^\mu)},
\eeq
and we recall $\delta(-x)=\delta(x)$.}
\beq
(g_0 k_u^2-\delta^{ij}k_i k_j)\tilde{G}(k)=i.
\eeq
The poles structure depends on the coupling constant $g_0$. Calling $k^2=\delta^{ij}k_ik_j$ and assuming $g_0$ positive from now on, we indeed have
$\tilde{G}(k)=\frac{i}{(\sqrt{g_0} k_u-\sqrt{k^2})(\sqrt{g_0} k_u+\sqrt{k^2})}$,
and the poles are located at $k_u=\pm \sqrt{\frac{k^2}{g_0}}$.
Choosing the time-ordered contour, that is, the $i\ve$-prescription $k_u=\pm \sqrt{\frac{k^2}{g_0}}\mp i\ve$, we get Feynman's Green function
\beq
G_F(x-x')=\int \frac{\text{d}^{d+1} k}{(2\pi)^{d+1}}\frac{ie^{-k_\mu(x^\mu-x'^\mu)}}{g_0 k_u^2-\delta^{ij}k_i k_j+i\ve}.
\eeq
This shows that the $u$-time-ordered two point function $G_F(x-x')=\langle 0 | T(\phi(x)\phi(x'))|0\rangle$ is also identical to the relativistic case, modulo the crucial replacement $c\leftrightarrow \frac1{\sqrt{g_0}}$. The dispersion relation $k_u=\pm \sqrt{\frac{k_ik_j\delta^{ij}}{g_0}}$ depends on the coupling constant. Therefore, wave packets are solutions of the Carrollian scalar field equation of motion, but their frequency is dictated by the coupling constant, defining the time order in this theory is thus subtle. Here we simply used the insertion of $\theta(u-u')$ to define Feynman's time-ordered Green function. A deeper study is required to understand the nature of this Carrollian motion, but from what we have achieved it seems that Carrollian waves move at different speeds, each determined by their coupling constant. Therefore, even when the energy is finite, motion is completely free, rather than completely frozen. 

Notice that in the relativistic vacuum we tune our variables such that relativistic waves have a dispersion relation which is given by $\omega=ck$. Generically, a wave propagating in a medium has a non-trivial dispersion relation. For instance, a string vibration has a modelled dispersion relation given by
\beq
\omega=\sqrt{\frac{T}{\mu}}k,
\eeq
where $\omega$ is the frequency, $k$ is the wavenumber, $T$ is the string tension and $\mu$ is the linear density of the string. 

Similarly, and even more suggestively, for a plane relativistic electromagnetic wave propagating in a uniform isotropic linear medium with electric permittivity $\epsilon$ and magnetic permeability $\mu$, the dispersion relation is
\beq
\omega=\frac{k}{\sqrt{\mu\epsilon}},
\eeq
To convert to our notation, we note that $k=\sqrt{k_ik_j\delta^{ij}}$ and $\omega=k_u$.
Then, comparing this equation with our dispersion relation $k_u=\pm \frac{k}{\sqrt{g_0}}$, we observe that it is as if a Carrollian wave propagates on a medium possessing a non-trivial electric permittivity and magnetic permeability such that 
\beq
g_0\approx \mu\epsilon.
\eeq
The main difference with the relativistic case is that each Carrollian scalar field propagates with a different $\mu$ and $\epsilon$. Then, each Carrollian scalar field would have a different dispersion in the Carrollian vacuum, a freer principle of relativity.
While this is merely an analogy at this point, it indicates a potentially interesting feature of the Carrollian vacuum. 

\section{Outlook}

We have shown that one can combine the electric and magnetic Carroll scalar action intrinsically on a Carrollian manifold. While in the $c\to 0$ limit these two actions cannot coexist, from the intrinsic viewpoint they naturally combine and lead to non-trivial dynamics. We discussed how the action behaves both under local Carroll boosts and diffeomorphisms. Both sectors are invariant under diffeomorphisms, but the magnetic part is not invariant under local Carroll boost, which might be a cause of concern, and deserve further study, although the possibility of this symmetry being (spontaneously) broken has been advocated in \cite{Armas:2023dcz}. 
This paper is a first exploration, and many refinements and generalizations can be pursued. Among them, one should perform this analysis on a general background, without requiring $h_{\mu\nu}$ to be time independent. Another aspect to explore is the symplectic analysis and canonical quantization. Indeed, although our results are classically sounded, there could be issues in the quantum theory, especially concerning the time ordering and the $i\ve$ prescription, since as we have seen the notion of frequency is coupling dependent. For instance, what are the properties of a Carrollian harmonic oscillator? Another urgent task is to compute how the coupling constant $g_0$ runs in the quantum theory. Other comments are in order at this stage.

Motion in Carroll physics has been already advocated from the hydrodynamics perspective \cite{Ciambelli:2018xat}. There however, we lacked a microscopic analysis, and one could have argued that it was a fictitious macroscopic evolution of thermodynamic quantities. The analysis performed here suggests that there is microscopic motion, and opens the door to explorations in  Carrollian kinetic theory. 

A similar avenue concerns the study of Carrollian algebras. The latter are typically introduced as contractions of the Poincar\'e algebra. But we have seen in this paper that intrinsic Carrollian physics might have sectors that are not reachable through a limiting procedure. Thus, it would be interesting to explore whether an intrinsic algebraic analysis can be done, such that the Hamiltonian does not commute with all generators, and thus motion is conceivable. 

In a companion paper \cite{Ciambelli:2023tzb}, we show that a strikingly similar analysis pertains Carrollian geodesics in Carroll gravity. While not reachable through a ultra-relativistic limit, there are intrinsic Carrollian geodesics showing that particles move on a Carrollian manifold. We thus believe that we have uncovered physical motion in Carrollian physics. This generalizes the already found motion in \cite{deBoer:2023fnj} for zero-energy particles and in \cite{Casalbuoni:2023bbh, Zhang:2023jbi} for multi-particle systems. Furthermore, it also generalizes the intrinsic analysis done in \cite{Gupta:2020dtl} for Carroll diffeomorphisms and in \cite{Bagchi:2022eav} in two spacetime dimensions.

Understanding the microscopic structure of Carrollian physics could have important repercussions in celestial holography, black hole physics and quantum gravity. In celestial holography, what we have found here is an intrinsic result in the bona fide boundary theory dual to asymptotically flat gravity. Can we access this dynamical field from the bulk, and if so how? Simultaneously, this might shed light on the physics at the black hole horizon. The membrane paradigm predicts that Einstein equations become Carrollian evolution equations at the horizon. Understanding the thermodynamic and microscopic properties in Carroll could thus help us understand the kinetic origin of black hole thermodynamics, and its repercussions in quantum gravity. Indeed, as already stated, this work opens the doors to the study of thermodynamics of Carrollian physics from a miscroscopic viewpoint, and thus Carrollian kinetic theory. 

In conclusion, there are many outlooks and future directions that we can pursue, related to all the Carrollian applications mentioned in the Introduction. This paper is just the starting point, and its aim is to show that Carrollian theories have still much to explore, especially from the intrinsic point of view.

\vspace{0.3cm}

\paragraph{Acknowledgements} This project started after the II Carroll workshop in Mons (Sep. 2022) and reached its completion after the III Carroll workshop in Thessaloniki (Oct. 2023). I thank the organizers and participants of both workshops for the stimulating environment. In particular, I am grateful to Stefano Baiguera and Gerben Oling for discussions at very early stages of this project. I have benefited from discussions with Luis Lehner, Andrea Puhm, and especially Sabrina Pasterski. I also wish to thank Ankit Aggarwal and Florian Ecker for important feedback on the manuscript. Furthermore, I am indebted with Daniel Grumiller for precious help and suggestions during the final stages of this project. I also thank Sruthi Narayanan, Marios Petropoulos, José Senovilla, Simone Speziale, and C\'eline Zwikel for discussions on Carrollian connections. Thank you Laurent Freidel, our long discussions have been essential, I greatly appreciate them. Research at Perimeter Institute is supported in part by the Government of Canada through the Department of Innovation, Science and Economic Development Canada and by the Province of Ontario through the Ministry of Colleges and Universities.

\appendix 

\section{Carrollian Connections\label{AppA}}

The goal of this Appendix is to show that for a time-independent degenerate metric $h_{\mu\nu}$ one can impose that the connection is both torsion-free and metric-compatible. For the sake of completeness, we  perform a full analysis of possible Carrollian connections. The latter have been studied in \cite{Duval:2014uva, Duval:2014uoa, Bekaert:2015xua, Ciambelli:2018xat, Ciambelli:2018wre, Morand:2018tke, Figueroa-OFarrill:2018ilb, Ciambelli:2019lap, Figueroa-OFarrill:2020gpr, Hansen:2021fxi, Chandrasekaran:2021hxc, Petkou:2022bmz, Freidel:2022vjq, Ciambelli:2023mir, Campoleoni:2023fug}.

\subsection{Warm up, Pseudo-Riemannian}

We start with a pseudo-Riemannian manifold $\cM$ as a warm up. The metric is a non-degenerate bilinear tensor, and we employ the notation
\beq
g:T\cM\otimes T\cM\to C^{\infty}(\cM)\quad g(X,Y)=g_{ab}X^a Y^b,
\eeq
for all $X,Y\in T\cM$ vector fields. The indices $a,b$ refer to an arbitrary basis $e_a$ of $T\cM$, such that
\beq
X=X^a e_a,\quad Y=Y^a e_a, \quad g(e_a,e_b)=g_{ab}.
\eeq
Typically, one considers a coordinate basis, or a specific non-coordinate basis where $g_{ab}=\eta_{ab}$, and thus a Lorentz bundle. We will consider an arbitrary frame, without particular restrictions on the G-structure.
The basis has thus generic structure constants
\beq
[e_a,e_b]=C_{ab}{}^c e_c,
\eeq
and for $C_{ab}{}^c=0$ we get the usual coordinates basis. We introduce an affine connection $\nabla_X:T\cM\to T\cM$. Affinity means, for all $f\in C^{\infty}(\cM)$,
\beq
\nabla_{fX}=f\nabla_X\qquad \nabla_X (fY)=f\nabla_X Y+X(f) Y,
\eeq
where $X(f)=X^a e_a(f)$. 

The key tensors are the torsion and the covariant derivative of the metric. The torsion is a skew symmetric bilinear tensor defined as
\beq\label{T}
T(X,Y)=\nabla_X Y-\nabla_Y X-[X,Y],
\eeq
where the commutator ensures  $T(fX,Y)=fT(X,Y)$. The covariant derivative of the metric is given by
\beq\label{dg}
\nabla_X g(Y,Z)=X(g(Y,Z))-g(\nabla_X Y,Z)-g(Y,\nabla_X Z).
\eeq
A connection satisfying $T(X,Y)=0$ for all $X,Y$ is called torsion-free, while the condition $\nabla_X g(Y,Z)=0$ for all $X,Y,Z$ is called metricity. The Christoffel symbols are defined by the action of the connection on the basis
\beq
\nabla_{e_a} e_b =\Gamma_{ab}^{c}e_c.
\eeq
The Levi-Civita result \cite{Levi-Civita:1917pgo} is that there exists a unique affine connection satisfying torsion-less and metricity. As usual, one solves these conditions (for the basis elements) and gets the Christoffel symbols
\beq
\Gamma_{ab}^c&=&\frac12 g^{cd}\Big(e_a(g_{bd})+e_b(g_{ad})-e_d(g_{ab}))\nn\\
&&+C_{ab}{}^{e} g_{ed}+C_{db}{}^{e} g_{ea}-C_{ad}{}^{e} g_{eb}\Big).
\eeq
The crucial observation is that the connection is entirely solved by the metric, and no conditions are imposed of the latter from setting \eqref{T} and \eqref{dg} to zero.

\subsection{Carrollian}

Consider now the ruled Carrollian structure discussed in Section \ref{s2}. It is convenient to go into a tangent bundle basis $e_a\in T\cM$ adapted to $v^\mu$. So we choose
\beq
e_a=e^\mu_a\pa_\mu=(e_0,e_A)= (v, e_A),
\eeq
where $a=(0,A)$ is a $d+1$-dimensional index, and by construction $e_0=v=v^\mu\pa_\mu$. The dual basis $e^a\in T^*\cM$ is defined via
\beq
e^a e_b=\delta^a_b,
\eeq
and thus
\beq
e^a=e_\mu^a \text{d}x^\mu=(E,e^A),
\eeq
where $E=E_\mu \text{d}x^\mu$ is the Ehresmann connection. The merit of this basis is to diagonalize the condition $v^\mu h_{\mu\nu}=0$. Indeed, defining the metric is this basis
\beq
q(e_a,e_b)=q_{ab},
\eeq
and using that $q_{ab}=e_a^\mu h_{\mu\nu}e^\nu_b$, we have
\beq
q(e_0,e_b)=q_{0b}=e_0^\mu h_{\mu\nu}e_b^\nu=v^\mu  h_{\mu\nu}e_b^\nu=0.
\eeq
Therefore, only $q_{AB}$ is non-vanishing.

We then introduce the affine connection in this basis
\beq
D_a e_b=D_{e_a} e_b=\Gamma_{ab}^c e_c.
\eeq
So far, the structure constants $C_{ab}{}^c$ are arbitrary. Following \cite{Ciambelli:2019lap}, given the underlying fibre bundle structure, we assume 
\beq
[v,e_A]=\varphi_A v,\qquad [e_A,e_B]=\omega_{AB} v,
\eeq
where $\varphi_A$ and $\omega_{AB}$ are known as the Carrollian acceleration and vorticity. This means that the only non-vanishing structure constants are
\beq
C_{0 A}{}^0=\varphi_A,\qquad C_{AB}{}^0=\omega_{AB}.
\eeq
One can easily generalize this to arbitrary $C_{ab}{}^c$, without changing the main result of this Appendix.

We then compute the torsion ($K_{[ab]}=\frac12 (K_{ab}-K_{ba})$)
\beq
T(v,v)&=&0\\
T(v,e_A)&=&(2\Gamma_{[0A]}^0-\varphi_A)v+2\Gamma_{[0A]}^Be_B\\
T(e_A,e_B)&=&(2\Gamma_{[AB]}^0-\omega_{AB})v +2\Gamma_{[AB]}^Ce_C.
\eeq
So far, the analysis is similar to the relativistic case, the main difference enters when discussing metricity, since we are dealing with a degenerate metric $q_{ab}$. Its covariant derivative is
\beq
&D_0 q_{00}=0\quad D_A q_{00}=0&\\
&D_0 q_{0A}=-\Gamma_{00}^Bq_{BA}\quad D_A q_{0B}=-\Gamma_{A0}^Cq_{CB}&\label{21}\\
&D_0 q_{AB}=v(q_{AB})-\Gamma_{0A}^Cq_{CB}-\Gamma_{0B}^Cq_{CA}&\label{22}\\
&D_A q_{BC}=e_A(q_{BC})-\Gamma_{AB}^Dq_{DC}-\Gamma_{AC}^Dq_{DB}\label{23},&
\eeq
where many simplifications occurred thanks to the choice of adapted basis, and we recall $D_0=D_{e_0}$ and $e_0=v$.

Now suppose we insist in requiring torsion-free and metricity. The torsion-free condition gives
\beq
&\Gamma_{[0A]}^0=\frac12 \varphi_A \quad \Gamma_{[0A]}^B=0 \quad \Gamma_{[AB]}^0=\frac12 \omega_{AB} \quad \Gamma_{[AB]}^C=0.& \nn
\eeq
For metricity, eq. \eqref{21} gives
\beq
\Gamma_{00}^B=0\quad \Gamma_{A0}^C=0,
\eeq
while eqs. \eqref{22} and \eqref{23} read
\beq
&v(q_{AB})=\Gamma_{0A}^Cq_{CB}+\Gamma_{0B}^Cq_{CA}&\\
&e_A(q_{BC})=\Gamma_{AB}^Dq_{DC}+\Gamma_{AC}^Dq_{DB}.&
\eeq
We immediately see that the condition $\Gamma_{[0A]}^B=0$ together with $\Gamma_{A0}^C=0$ implies
\beq
v(q_{AB})=0.
\eeq
This is the main message of this Appendix. Torsion-free and metricity on a generic Carrollian structure imply that the degenerate metric is time independent. This is different from the pseudo-Riemannian case, where these two conditions have no restrictions on the underlying geometry. There are two options, either we are on a time-independent background, and then we can select a torsion-free and metric compatible connection $D$, or we need to give up one of these conditions. In the main body of the manuscript, we decided to work on a time-independent background, for the three reasons explained below \eqref{bq}. 

We note that torsion-free or metricity can be separately imposed without leading to impositions on the background. Torsion-free is imposed in \cite{Ciambelli:2023mir}, and then there is some intrinsic leftover non-metricity. Other accounts, such as \cite{Figueroa-OFarrill:2020gpr}, impose metricity, and then there is a leftover intrinsic torsion. Depending on the system under scrutiny, a choice could be more helpful than the other. For instance, since the seminal work of Ashtekar \cite{Ashtekar:1981hw}, radiation in an asymptotically flat spacetime is conveniently included in the connection, and thus the torsion-free condition is preferred. Conversely, it is clearly advantageous to keep torsion but impose metricity when discussing Carrollian fermions. The important result to retain is that there is no unique choice for a Carrollian connection, and that imposing too stringent conditions leads to constraints on the geometric background.

\bibliographystyle{uiuchept}
\bibliography{ArxivV2.bib}

\providecommand{\href}[2]{#2}\begingroup\raggedright\begin{thebibliography}{10}

\bibitem{Levy1965}
J.-M. Lévy-Leblond, ``Une nouvelle limite non-relativiste du groupe de
  poincaré,'' {\em Annales de l'I.H.P. Physique théorique} {\bf 3} (1965)
  no.~1, 1--12. \url{http://eudml.org/doc/75509}.

\bibitem{Gupta1966}
N.~D. SenGupta, ``On an analogue of the galilei group,''
  \href{http://dx.doi.org/10.1007/bf02740871}{{\em Il Nuovo Cimento A Series
  10} {\bf 44} (1966) no.~2, 512--517}.

\bibitem{Duval:2014lpa}
C.~Duval, G.~W. Gibbons, and P.~A. Horvathy, ``{Conformal Carroll groups},''
  \href{http://dx.doi.org/10.1088/1751-8113/47/33/335204}{{\em J. Phys. A} {\bf
  47} (2014) no.~33, 335204}, \href{http://arxiv.org/abs/1403.4213}{{\tt
  arXiv:1403.4213 [hep-th]}}.

\bibitem{Duval:2014uva}
C.~Duval, G.~W. Gibbons, and P.~A. Horvathy, ``{Conformal Carroll groups and
  BMS symmetry},'' \href{http://dx.doi.org/10.1088/0264-9381/31/9/092001}{{\em
  Class. Quant. Grav.} {\bf 31} (2014)  092001},
  \href{http://arxiv.org/abs/1402.5894}{{\tt arXiv:1402.5894 [gr-qc]}}.

\bibitem{Duval:2014uoa}
C.~Duval, G.~W. Gibbons, P.~A. Horvathy, and P.~M. Zhang, ``{Carroll versus
  Newton and Galilei: two dual non-Einsteinian concepts of time},''
  \href{http://dx.doi.org/10.1088/0264-9381/31/8/085016}{{\em Class. Quant.
  Grav.} {\bf 31} (2014)  085016}, \href{http://arxiv.org/abs/1402.0657}{{\tt
  arXiv:1402.0657 [gr-qc]}}.

\bibitem{Ciambelli:2019lap}
L.~Ciambelli, R.~G. Leigh, C.~Marteau, and P.~M. Petropoulos, ``{Carroll
  Structures, Null Geometry and Conformal Isometries},''
  \href{http://dx.doi.org/10.1103/PhysRevD.100.046010}{{\em Phys. Rev. D} {\bf
  100} (2019) no.~4, 046010}, \href{http://arxiv.org/abs/1905.02221}{{\tt
  arXiv:1905.02221 [hep-th]}}.

\bibitem{deBoer:2017ing}
J.~de~Boer, J.~Hartong, N.~A. Obers, W.~Sybesma, and S.~Vandoren, ``{Perfect
  Fluids},'' \href{http://dx.doi.org/10.21468/SciPostPhys.5.1.003}{{\em SciPost
  Phys.} {\bf 5} (2018) no.~1, 003},
  \href{http://arxiv.org/abs/1710.04708}{{\tt arXiv:1710.04708 [hep-th]}}.

\bibitem{Ciambelli:2018xat}
L.~Ciambelli, C.~Marteau, A.~C. Petkou, P.~M. Petropoulos, and K.~Siampos,
  ``{Covariant Galilean versus Carrollian hydrodynamics from relativistic
  fluids},'' \href{http://dx.doi.org/10.1088/1361-6382/aacf1a}{{\em Class.
  Quant. Grav.} {\bf 35} (2018) no.~16, 165001},
  \href{http://arxiv.org/abs/1802.05286}{{\tt arXiv:1802.05286 [hep-th]}}.

\bibitem{Ciambelli:2018ojf}
L.~Ciambelli and C.~Marteau, ``{Carrollian conservation laws and Ricci-flat
  gravity},'' \href{http://dx.doi.org/10.1088/1361-6382/ab0d37}{{\em Class.
  Quant. Grav.} {\bf 36} (2019) no.~8, 085004},
  \href{http://arxiv.org/abs/1810.11037}{{\tt arXiv:1810.11037 [hep-th]}}.

\bibitem{Campoleoni:2018ltl}
A.~Campoleoni, L.~Ciambelli, C.~Marteau, P.~M. Petropoulos, and K.~Siampos,
  ``{Two-dimensional fluids and their holographic duals},''
  \href{http://dx.doi.org/10.1016/j.nuclphysb.2019.114692}{{\em Nucl. Phys. B}
  {\bf 946} (2019)  114692}, \href{http://arxiv.org/abs/1812.04019}{{\tt
  arXiv:1812.04019 [hep-th]}}.

\bibitem{Ciambelli:2020eba}
L.~Ciambelli, C.~Marteau, P.~M. Petropoulos, and R.~Ruzziconi, ``{Gauges in
  Three-Dimensional Gravity and Holographic Fluids},''
  \href{http://dx.doi.org/10.1007/JHEP11(2020)092}{{\em JHEP} {\bf 11} (2020)
  092}, \href{http://arxiv.org/abs/2006.10082}{{\tt arXiv:2006.10082
  [hep-th]}}.

\bibitem{Ciambelli:2020ftk}
L.~Ciambelli, C.~Marteau, P.~M. Petropoulos, and R.~Ruzziconi,
  ``{Fefferman-Graham and Bondi Gauges in the Fluid/Gravity Correspondence},''
  \href{http://dx.doi.org/10.22323/1.376.0154}{{\em PoS} {\bf CORFU2019} (2020)
   154}, \href{http://arxiv.org/abs/2006.10083}{{\tt arXiv:2006.10083
  [hep-th]}}.

\bibitem{Freidel:2022bai}
L.~Freidel and P.~Jai-akson, ``{Carrollian hydrodynamics from symmetries},''
  \href{http://dx.doi.org/10.1088/1361-6382/acb194}{{\em Class. Quant. Grav.}
  {\bf 40} (2023) no.~5, 055009}, \href{http://arxiv.org/abs/2209.03328}{{\tt
  arXiv:2209.03328 [hep-th]}}.

\bibitem{Petkou:2022bmz}
A.~C. Petkou, P.~M. Petropoulos, D.~R. Betancour, and K.~Siampos,
  ``{Relativistic fluids, hydrodynamic frames and their Galilean versus
  Carrollian avatars},'' \href{http://dx.doi.org/10.1007/JHEP09(2022)162}{{\em
  JHEP} {\bf 09} (2022)  162}, \href{http://arxiv.org/abs/2205.09142}{{\tt
  arXiv:2205.09142 [hep-th]}}.

\bibitem{Bagchi:2016bcd}
A.~Bagchi, R.~Basu, A.~Kakkar, and A.~Mehra, ``{Flat Holography: Aspects of the
  dual field theory},'' \href{http://dx.doi.org/10.1007/JHEP12(2016)147}{{\em
  JHEP} {\bf 12} (2016)  147}, \href{http://arxiv.org/abs/1609.06203}{{\tt
  arXiv:1609.06203 [hep-th]}}.

\bibitem{Ciambelli:2018wre}
L.~Ciambelli, C.~Marteau, A.~C. Petkou, P.~M. Petropoulos, and K.~Siampos,
  ``{Flat holography and Carrollian fluids},''
  \href{http://dx.doi.org/10.1007/JHEP07(2018)165}{{\em JHEP} {\bf 07} (2018)
  165}, \href{http://arxiv.org/abs/1802.06809}{{\tt arXiv:1802.06809
  [hep-th]}}.

\bibitem{Bagchi:2019xfx}
A.~Bagchi, A.~Mehra, and P.~Nandi, ``{Field Theories with Conformal Carrollian
  Symmetry},'' \href{http://dx.doi.org/10.1007/JHEP05(2019)108}{{\em JHEP} {\bf
  05} (2019)  108}, \href{http://arxiv.org/abs/1901.10147}{{\tt
  arXiv:1901.10147 [hep-th]}}.

\bibitem{Bagchi:2019clu}
A.~Bagchi, R.~Basu, A.~Mehra, and P.~Nandi, ``{Field Theories on Null
  Manifolds},'' \href{http://dx.doi.org/10.1007/JHEP02(2020)141}{{\em JHEP}
  {\bf 02} (2020)  141}, \href{http://arxiv.org/abs/1912.09388}{{\tt
  arXiv:1912.09388 [hep-th]}}.

\bibitem{Gupta:2020dtl}
N.~Gupta and N.~V. Suryanarayana, ``{Constructing Carrollian CFTs},''
  \href{http://dx.doi.org/10.1007/JHEP03(2021)194}{{\em JHEP} {\bf 03} (2021)
  194}, \href{http://arxiv.org/abs/2001.03056}{{\tt arXiv:2001.03056
  [hep-th]}}.

\bibitem{Banerjee:2020qjj}
K.~Banerjee, R.~Basu, A.~Mehra, A.~Mohan, and A.~Sharma, ``{Interacting
  Conformal Carrollian Theories: Cues from Electrodynamics},''
  \href{http://dx.doi.org/10.1103/PhysRevD.103.105001}{{\em Phys. Rev. D} {\bf
  103} (2021) no.~10, 105001}, \href{http://arxiv.org/abs/2008.02829}{{\tt
  arXiv:2008.02829 [hep-th]}}.

\bibitem{Figueroa-OFarrill:2021sxz}
J.~Figueroa-O'Farrill, E.~Have, S.~Prohazka, and J.~Salzer, ``{Carrollian and
  celestial spaces at infinity},''
  \href{http://dx.doi.org/10.1007/JHEP09(2022)007}{{\em JHEP} {\bf 09} (2022)
  007}, \href{http://arxiv.org/abs/2112.03319}{{\tt arXiv:2112.03319
  [hep-th]}}.

\bibitem{Chen:2021xkw}
B.~Chen, R.~Liu, and Y.-f. Zheng, ``{On higher-dimensional Carrollian and
  Galilean conformal field theories},''
  \href{http://dx.doi.org/10.21468/SciPostPhys.14.5.088}{{\em SciPost Phys.}
  {\bf 14} (2023) no.~5, 088}, \href{http://arxiv.org/abs/2112.10514}{{\tt
  arXiv:2112.10514 [hep-th]}}.

\bibitem{Herfray:2021qmp}
Y.~Herfray, ``{Carrollian manifolds and null infinity: a view from Cartan
  geometry},'' \href{http://dx.doi.org/10.1088/1361-6382/ac635f}{{\em Class.
  Quant. Grav.} {\bf 39} (2022) no.~21, 215005},
  \href{http://arxiv.org/abs/2112.09048}{{\tt arXiv:2112.09048 [gr-qc]}}.

\bibitem{Donnay:2022aba}
L.~Donnay, A.~Fiorucci, Y.~Herfray, and R.~Ruzziconi, ``{Carrollian Perspective
  on Celestial Holography},''
  \href{http://dx.doi.org/10.1103/PhysRevLett.129.071602}{{\em Phys. Rev.
  Lett.} {\bf 129} (2022) no.~7, 071602},
  \href{http://arxiv.org/abs/2202.04702}{{\tt arXiv:2202.04702 [hep-th]}}.

\bibitem{Bagchi:2022eav}
A.~Bagchi, A.~Banerjee, S.~Dutta, K.~S. Kolekar, and P.~Sharma, ``{Carroll
  covariant scalar fields in two dimensions},''
  \href{http://dx.doi.org/10.1007/JHEP01(2023)072}{{\em JHEP} {\bf 01} (2023)
  072}, \href{http://arxiv.org/abs/2203.13197}{{\tt arXiv:2203.13197
  [hep-th]}}.

\bibitem{Bekaert:2022oeh}
X.~Bekaert, A.~Campoleoni, and S.~Pekar, ``{Carrollian conformal scalar as
  flat-space singleton},''
  \href{http://dx.doi.org/10.1016/j.physletb.2023.137734}{{\em Phys. Lett. B}
  {\bf 838} (2023)  137734}, \href{http://arxiv.org/abs/2211.16498}{{\tt
  arXiv:2211.16498 [hep-th]}}.

\bibitem{Bagchi:2022emh}
A.~Bagchi, S.~Banerjee, R.~Basu, and S.~Dutta, ``{Scattering Amplitudes:
  Celestial and Carrollian},''
  \href{http://dx.doi.org/10.1103/PhysRevLett.128.241601}{{\em Phys. Rev.
  Lett.} {\bf 128} (2022) no.~24, 241601},
  \href{http://arxiv.org/abs/2202.08438}{{\tt arXiv:2202.08438 [hep-th]}}.

\bibitem{Rivera-Betancour:2022lkc}
D.~Rivera-Betancour and M.~Vilatte, ``{Revisiting the Carrollian scalar
  field},'' \href{http://dx.doi.org/10.1103/PhysRevD.106.085004}{{\em Phys.
  Rev. D} {\bf 106} (2022) no.~8, 085004},
  \href{http://arxiv.org/abs/2207.01647}{{\tt arXiv:2207.01647 [hep-th]}}.

\bibitem{Baiguera:2022lsw}
S.~Baiguera, G.~Oling, W.~Sybesma, and B.~T. S\o{}gaard, ``{Conformal Carroll
  scalars with boosts},''
  \href{http://dx.doi.org/10.21468/SciPostPhys.14.4.086}{{\em SciPost Phys.}
  {\bf 14} (2023) no.~4, 086}, \href{http://arxiv.org/abs/2207.03468}{{\tt
  arXiv:2207.03468 [hep-th]}}.

\bibitem{Campoleoni:2022wmf}
A.~Campoleoni, L.~Ciambelli, A.~Delfante, C.~Marteau, P.~M. Petropoulos, and
  R.~Ruzziconi, ``{Holographic Lorentz and Carroll frames},''
  \href{http://dx.doi.org/10.1007/JHEP12(2022)007}{{\em JHEP} {\bf 12} (2022)
  007}, \href{http://arxiv.org/abs/2208.07575}{{\tt arXiv:2208.07575
  [hep-th]}}.

\bibitem{Donnay:2022wvx}
L.~Donnay, A.~Fiorucci, Y.~Herfray, and R.~Ruzziconi, ``{Bridging Carrollian
  and celestial holography},''
  \href{http://dx.doi.org/10.1103/PhysRevD.107.126027}{{\em Phys. Rev. D} {\bf
  107} (2023) no.~12, 126027}, \href{http://arxiv.org/abs/2212.12553}{{\tt
  arXiv:2212.12553 [hep-th]}}.

\bibitem{Dutta:2022vkg}
S.~Dutta, ``{Stress tensors of 3d Carroll CFTs},''
  \href{http://arxiv.org/abs/2212.11002}{{\tt arXiv:2212.11002 [hep-th]}}.

\bibitem{Mittal:2022ywl}
N.~Mittal, P.~M. Petropoulos, D.~Rivera-Betancour, and M.~Vilatte, ``{Ehlers,
  Carroll, charges and dual charges},''
  \href{http://dx.doi.org/10.1007/JHEP07(2023)065}{{\em JHEP} {\bf 07} (2023)
  065}, \href{http://arxiv.org/abs/2212.14062}{{\tt arXiv:2212.14062
  [hep-th]}}.

\bibitem{Chen:2023pqf}
B.~Chen, R.~Liu, H.~Sun, and Y.-f. Zheng, ``{Constructing Carrollian Field
  Theories from Null Reduction},'' \href{http://arxiv.org/abs/2301.06011}{{\tt
  arXiv:2301.06011 [hep-th]}}.

\bibitem{Mehra:2023rmm}
A.~Mehra and A.~Sharma, ``{Toward Carrollian quantization: Renormalization of
  Carrollian electrodynamics},''
  \href{http://dx.doi.org/10.1103/PhysRevD.108.046019}{{\em Phys. Rev. D} {\bf
  108} (2023) no.~4, 046019}, \href{http://arxiv.org/abs/2302.13257}{{\tt
  arXiv:2302.13257 [hep-th]}}.

\bibitem{Campoleoni:2023fug}
A.~Campoleoni, A.~Delfante, S.~Pekar, P.~M. Petropoulos, D.~Rivera-Betancour,
  and M.~Vilatte, ``{Flat from anti-de Sitter},''
  \href{http://arxiv.org/abs/2309.15182}{{\tt arXiv:2309.15182 [hep-th]}}.

\bibitem{Salzer:2023jqv}
J.~Salzer, ``{An embedding space approach to Carrollian CFT correlators for
  flat space holography},''
  \href{http://dx.doi.org/10.1007/JHEP10(2023)084}{{\em JHEP} {\bf 10} (2023)
  084}, \href{http://arxiv.org/abs/2304.08292}{{\tt arXiv:2304.08292
  [hep-th]}}.

\bibitem{Bagchi:2023fbj}
A.~Bagchi, P.~Dhivakar, and S.~Dutta, ``{AdS Witten diagrams to Carrollian
  correlators},'' \href{http://dx.doi.org/10.1007/JHEP04(2023)135}{{\em JHEP}
  {\bf 04} (2023)  135}, \href{http://arxiv.org/abs/2303.07388}{{\tt
  arXiv:2303.07388 [hep-th]}}.

\bibitem{Saha:2023hsl}
A.~Saha, ``{Carrollian approach to 1 + 3D flat holography},''
  \href{http://dx.doi.org/10.1007/JHEP06(2023)051}{{\em JHEP} {\bf 06} (2023)
  051}, \href{http://arxiv.org/abs/2304.02696}{{\tt arXiv:2304.02696
  [hep-th]}}.

\bibitem{Banerjee:2023jpi}
K.~Banerjee, R.~Basu, B.~Krishnan, S.~Maulik, A.~Mehra, and A.~Ray, ``{One-loop
  quantum effects in Carroll scalars},''
  \href{http://dx.doi.org/10.1103/PhysRevD.108.085022}{{\em Phys. Rev. D} {\bf
  108} (2023) no.~8, 085022}, \href{http://arxiv.org/abs/2307.03901}{{\tt
  arXiv:2307.03901 [hep-th]}}.

\bibitem{Penna:2015gza}
R.~F. Penna, ``{BMS invariance and the membrane paradigm},''
  \href{http://dx.doi.org/10.1007/JHEP03(2016)023}{{\em JHEP} {\bf 03} (2016)
  023}, \href{http://arxiv.org/abs/1508.06577}{{\tt arXiv:1508.06577
  [hep-th]}}.

\bibitem{Penna:2018gfx}
R.~F. Penna, ``{Near-horizon Carroll symmetry and black hole Love numbers},''
  \href{http://arxiv.org/abs/1812.05643}{{\tt arXiv:1812.05643 [hep-th]}}.

\bibitem{Donnay:2019jiz}
L.~Donnay and C.~Marteau, ``{Carrollian Physics at the Black Hole Horizon},''
  \href{http://dx.doi.org/10.1088/1361-6382/ab2fd5}{{\em Class. Quant. Grav.}
  {\bf 36} (2019) no.~16, 165002}, \href{http://arxiv.org/abs/1903.09654}{{\tt
  arXiv:1903.09654 [hep-th]}}.

\bibitem{Redondo-Yuste:2022czg}
J.~Redondo-Yuste and L.~Lehner, ``{Non-linear black hole dynamics and
  Carrollian fluids},'' \href{http://dx.doi.org/10.1007/JHEP02(2023)240}{{\em
  JHEP} {\bf 02} (2023)  240}, \href{http://arxiv.org/abs/2212.06175}{{\tt
  arXiv:2212.06175 [gr-qc]}}.

\bibitem{Freidel:2022vjq}
L.~Freidel and P.~Jai-akson, ``{Carrollian hydrodynamics and symplectic
  structure on stretched horizons},''
  \href{http://arxiv.org/abs/2211.06415}{{\tt arXiv:2211.06415 [gr-qc]}}.

\bibitem{Gray:2022svz}
F.~Gray, D.~Kubiznak, T.~R. Perche, and J.~Redondo-Yuste, ``{Carrollian motion
  in magnetized black hole horizons},''
  \href{http://dx.doi.org/10.1103/PhysRevD.107.064009}{{\em Phys. Rev. D} {\bf
  107} (2023) no.~6, 064009}, \href{http://arxiv.org/abs/2211.13695}{{\tt
  arXiv:2211.13695 [gr-qc]}}.

\bibitem{Ciambelli:2023mir}
L.~Ciambelli, L.~Freidel, and R.~G. Leigh, ``{Null Raychaudhuri: Canonical
  Structure and the Dressing Time},''
  \href{http://arxiv.org/abs/2309.03932}{{\tt arXiv:2309.03932 [hep-th]}}.

\bibitem{Ciambelli:2023mvj}
L.~Ciambelli and L.~Lehner, ``{Fluid-gravity correspondence and causal
  first-order relativistic viscous hydrodynamics},''
  \href{http://arxiv.org/abs/2310.15427}{{\tt arXiv:2310.15427 [hep-th]}}.

\bibitem{Bergshoeff:2014jla}
E.~Bergshoeff, J.~Gomis, and G.~Longhi, ``{Dynamics of Carroll Particles},''
  \href{http://dx.doi.org/10.1088/0264-9381/31/20/205009}{{\em Class. Quant.
  Grav.} {\bf 31} (2014) no.~20, 205009},
  \href{http://arxiv.org/abs/1405.2264}{{\tt arXiv:1405.2264 [hep-th]}}.

\bibitem{Hartong:2015xda}
J.~Hartong, ``{Gauging the Carroll Algebra and Ultra-Relativistic Gravity},''
  \href{http://dx.doi.org/10.1007/JHEP08(2015)069}{{\em JHEP} {\bf 08} (2015)
  069}, \href{http://arxiv.org/abs/1505.05011}{{\tt arXiv:1505.05011
  [hep-th]}}.

\bibitem{Henneaux:2021yzg}
M.~Henneaux and P.~Salgado-Rebolledo, ``{Carroll contractions of
  Lorentz-invariant theories},''
  \href{http://dx.doi.org/10.1007/JHEP11(2021)180}{{\em JHEP} {\bf 11} (2021)
  180}, \href{http://arxiv.org/abs/2109.06708}{{\tt arXiv:2109.06708
  [hep-th]}}.

\bibitem{Marsot:2022imf}
L.~Marsot, P.~M. Zhang, M.~Chernodub, and P.~A. Horvathy, ``{Hall effects in
  Carroll dynamics},''
  \href{http://dx.doi.org/10.1016/j.physrep.2023.07.007}{{\em Phys. Rept.} {\bf
  1028} (2023)  1--60}, \href{http://arxiv.org/abs/2212.02360}{{\tt
  arXiv:2212.02360 [hep-th]}}.

\bibitem{Bergshoeff:2022eog}
E.~Bergshoeff, J.~Figueroa-O'Farrill, and J.~Gomis, ``{A non-lorentzian
  primer},'' \href{http://dx.doi.org/10.21468/SciPostPhysLectNotes.69}{{\em
  SciPost Phys. Lect. Notes} {\bf 69} (2023)  1},
  \href{http://arxiv.org/abs/2206.12177}{{\tt arXiv:2206.12177 [hep-th]}}.

\bibitem{Bagchi:2022eui}
A.~Bagchi, A.~Banerjee, R.~Basu, M.~Islam, and S.~Mondal, ``{Magic fermions:
  Carroll and flat bands},''
  \href{http://dx.doi.org/10.1007/JHEP03(2023)227}{{\em JHEP} {\bf 03} (2023)
  227}, \href{http://arxiv.org/abs/2211.11640}{{\tt arXiv:2211.11640
  [hep-th]}}.

\bibitem{Kasikci:2023tvs}
O.~Kasikci, M.~Ozkan, and Y.~Pang, ``{Carrollian origin of spacetime subsystem
  symmetry},'' \href{http://dx.doi.org/10.1103/PhysRevD.108.045020}{{\em Phys.
  Rev. D} {\bf 108} (2023) no.~4, 045020},
  \href{http://arxiv.org/abs/2304.11331}{{\tt arXiv:2304.11331 [hep-th]}}.

\bibitem{Casalbuoni:2023bbh}
R.~Casalbuoni, D.~Dominici, and J.~Gomis, ``{Two interacting conformal Carroll
  particles},'' \href{http://dx.doi.org/10.1103/PhysRevD.108.086005}{{\em Phys.
  Rev. D} {\bf 108} (2023) no.~8, 086005},
  \href{http://arxiv.org/abs/2306.02614}{{\tt arXiv:2306.02614 [hep-th]}}.

\bibitem{Cerdeira:2023ztm}
J.~L.~V. Cerdeira, J.~Gomis, and A.~Kleinschmidt, ``{Non-Lorentzian expansions
  of the Lorentz force and kinematical algebras},''
  \href{http://arxiv.org/abs/2310.15245}{{\tt arXiv:2310.15245 [hep-th]}}.

\bibitem{Zhang:2023jbi}
P.~M. Zhang, H.-X. Zeng, and P.~A. Horvathy, ``{MultiCarroll dynamics},''
  \href{http://arxiv.org/abs/2306.07002}{{\tt arXiv:2306.07002 [gr-qc]}}.

\bibitem{Kamenshchik:2023kxi}
A.~Kamenshchik and F.~Muscolino, ``{Looking for Carroll particles in two time
  spacetime},'' \href{http://arxiv.org/abs/2310.19050}{{\tt arXiv:2310.19050
  [hep-th]}}.

\bibitem{Henneaux:1979vn}
M.~Henneaux, ``{Geometry of Zero Signature Space-times},'' {\em Bull. Soc.
  Math. Belg.} {\bf 31} (1979)  47--63.

\bibitem{Bergshoeff:2017btm}
E.~Bergshoeff, J.~Gomis, B.~Rollier, J.~Rosseel, and T.~ter Veldhuis,
  ``{Carroll versus Galilei Gravity},''
  \href{http://dx.doi.org/10.1007/JHEP03(2017)165}{{\em JHEP} {\bf 03} (2017)
  165}, \href{http://arxiv.org/abs/1701.06156}{{\tt arXiv:1701.06156
  [hep-th]}}.

\bibitem{Matulich:2019cdo}
J.~Matulich, S.~Prohazka, and J.~Salzer, ``{Limits of three-dimensional gravity
  and metric kinematical Lie algebras in any dimension},''
  \href{http://dx.doi.org/10.1007/JHEP07(2019)118}{{\em JHEP} {\bf 07} (2019)
  118}, \href{http://arxiv.org/abs/1903.09165}{{\tt arXiv:1903.09165
  [hep-th]}}.

\bibitem{Gomis:2020wxp}
J.~Gomis, D.~Hidalgo, and P.~Salgado-Rebolledo, ``{Non-relativistic and
  Carrollian limits of Jackiw-Teitelboim gravity},''
  \href{http://dx.doi.org/10.1007/JHEP05(2021)162}{{\em JHEP} {\bf 05} (2021)
  162}, \href{http://arxiv.org/abs/2011.15053}{{\tt arXiv:2011.15053
  [hep-th]}}.

\bibitem{Grumiller:2020elf}
D.~Grumiller, J.~Hartong, S.~Prohazka, and J.~Salzer, ``{Limits of JT
  gravity},'' \href{http://dx.doi.org/10.1007/JHEP02(2021)134}{{\em JHEP} {\bf
  02} (2021)  134}, \href{http://arxiv.org/abs/2011.13870}{{\tt
  arXiv:2011.13870 [hep-th]}}.

\bibitem{Concha:2021jnn}
P.~Concha, D.~Pe\~nafiel, L.~Ravera, and E.~Rodr\'\i{}guez,
  ``{Three-dimensional Maxwellian Carroll gravity theory and the cosmological
  constant},'' \href{http://dx.doi.org/10.1016/j.physletb.2021.136735}{{\em
  Phys. Lett. B} {\bf 823} (2021)  136735},
  \href{http://arxiv.org/abs/2107.05716}{{\tt arXiv:2107.05716 [hep-th]}}.

\bibitem{Perez:2021abf}
A.~P\'erez, ``{Asymptotic symmetries in Carrollian theories of gravity},''
  \href{http://dx.doi.org/10.1007/JHEP12(2021)173}{{\em JHEP} {\bf 12} (2021)
  173}, \href{http://arxiv.org/abs/2110.15834}{{\tt arXiv:2110.15834
  [hep-th]}}.

\bibitem{deBoer:2021jej}
J.~de~Boer, J.~Hartong, N.~A. Obers, W.~Sybesma, and S.~Vandoren, ``{Carroll
  Symmetry, Dark Energy and Inflation},''
  \href{http://dx.doi.org/10.3389/fphy.2022.810405}{{\em Front. in Phys.} {\bf
  10} (2022)  810405}, \href{http://arxiv.org/abs/2110.02319}{{\tt
  arXiv:2110.02319 [hep-th]}}.

\bibitem{Hansen:2021fxi}
D.~Hansen, N.~A. Obers, G.~Oling, and B.~T. S\o{}gaard, ``{Carroll Expansion of
  General Relativity},''
  \href{http://dx.doi.org/10.21468/SciPostPhys.13.3.055}{{\em SciPost Phys.}
  {\bf 13} (2022) no.~3, 055}, \href{http://arxiv.org/abs/2112.12684}{{\tt
  arXiv:2112.12684 [hep-th]}}.

\bibitem{Campoleoni:2022ebj}
A.~Campoleoni, M.~Henneaux, S.~Pekar, A.~P\'erez, and P.~Salgado-Rebolledo,
  ``{Magnetic Carrollian gravity from the Carroll algebra},''
  \href{http://dx.doi.org/10.1007/JHEP09(2022)127}{{\em JHEP} {\bf 09} (2022)
  127}, \href{http://arxiv.org/abs/2207.14167}{{\tt arXiv:2207.14167
  [hep-th]}}.

\bibitem{Ekiz:2022wbi}
E.~Ekiz, O.~Kasikci, M.~Ozkan, C.~B. Senisik, and U.~Zorba, ``{Non-relativistic
  and ultra-relativistic scaling limits of multimetric gravity},''
  \href{http://dx.doi.org/10.1007/JHEP10(2022)151}{{\em JHEP} {\bf 10} (2022)
  151}, \href{http://arxiv.org/abs/2207.07882}{{\tt arXiv:2207.07882
  [hep-th]}}.

\bibitem{Figueroa-OFarrill:2022mcy}
J.~Figueroa-O'Farrill, E.~Have, S.~Prohazka, and J.~Salzer, ``{The gauging
  procedure and carrollian gravity},''
  \href{http://dx.doi.org/10.1007/JHEP09(2022)243}{{\em JHEP} {\bf 09} (2022)
  243}, \href{http://arxiv.org/abs/2206.14178}{{\tt arXiv:2206.14178
  [hep-th]}}.

\bibitem{Miskovic:2023zfz}
O.~Miskovic, R.~Olea, P.~M. Petropoulos, D.~Rivera-Betancour, and K.~Siampos,
  ``{Chern-Simons action and the Carrollian Cotton tensors},''
  \href{http://arxiv.org/abs/2310.19929}{{\tt arXiv:2310.19929 [hep-th]}}.

\bibitem{Ecker:2023uwm}
F.~Ecker, D.~Grumiller, J.~Hartong, A.~P\'erez, S.~Prohazka, and R.~Troncoso,
  ``{Carroll black holes},'' \href{http://arxiv.org/abs/2308.10947}{{\tt
  arXiv:2308.10947 [hep-th]}}.

\bibitem{Ravera:2019ize}
L.~Ravera, ``{AdS Carroll Chern-Simons supergravity in 2 + 1 dimensions and its
  flat limit},'' \href{http://dx.doi.org/10.1016/j.physletb.2019.06.026}{{\em
  Phys. Lett. B} {\bf 795} (2019)  331--338},
  \href{http://arxiv.org/abs/1905.00766}{{\tt arXiv:1905.00766 [hep-th]}}.

\bibitem{Ali:2019jjp}
F.~Ali and L.~Ravera, ``{$\mathcal{N}$-extended Chern-Simons Carrollian
  supergravities in $2+1$ spacetime dimensions},''
  \href{http://dx.doi.org/10.1007/JHEP02(2020)128}{{\em JHEP} {\bf 02} (2020)
  128}, \href{http://arxiv.org/abs/1912.04172}{{\tt arXiv:1912.04172
  [hep-th]}}.

\bibitem{Ravera:2022buz}
L.~Ravera and U.~Zorba, ``{Carrollian and non-relativistic
  Jackiw\textendash{}Teitelboim supergravity},''
  \href{http://dx.doi.org/10.1140/epjc/s10052-023-11239-x}{{\em Eur. Phys. J.
  C} {\bf 83} (2023) no.~2, 107}, \href{http://arxiv.org/abs/2204.09643}{{\tt
  arXiv:2204.09643 [hep-th]}}.

\bibitem{Kasikci:2023zdn}
O.~Kasikci, M.~Ozkan, Y.~Pang, and U.~Zorba, ``{Carrollian Supersymmetry and
  SYK-like models},'' \href{http://arxiv.org/abs/2311.00039}{{\tt
  arXiv:2311.00039 [hep-th]}}.

\bibitem{Schild:1976vq}
A.~Schild, ``{Classical Null Strings},''
  \href{http://dx.doi.org/10.1103/PhysRevD.16.1722}{{\em Phys. Rev. D} {\bf 16}
  (1977)  1722}.

\bibitem{Isberg:1993av}
J.~Isberg, U.~Lindstrom, B.~Sundborg, and G.~Theodoridis, ``{Classical and
  quantized tensionless strings},''
  \href{http://dx.doi.org/10.1016/0550-3213(94)90056-6}{{\em Nucl. Phys. B}
  {\bf 411} (1994)  122--156}, \href{http://arxiv.org/abs/hep-th/9307108}{{\tt
  arXiv:hep-th/9307108}}.

\bibitem{Bagchi:2015nca}
A.~Bagchi, S.~Chakrabortty, and P.~Parekh, ``{Tensionless Strings from
  Worldsheet Symmetries},''
  \href{http://dx.doi.org/10.1007/JHEP01(2016)158}{{\em JHEP} {\bf 01} (2016)
  158}, \href{http://arxiv.org/abs/1507.04361}{{\tt arXiv:1507.04361
  [hep-th]}}.

\bibitem{Fursaev:2023lxq}
D.~V. Fursaev and I.~G. Pirozhenko, ``{Electromagnetic Waves from Pulsars
  Generated by Null Cosmic Strings},''
  \href{http://arxiv.org/abs/2309.01272}{{\tt arXiv:2309.01272 [gr-qc]}}.

\bibitem{Fursaev:2023oep}
D.~V. Fursaev, E.~A. Davydov, I.~G. Pirozhenko, and V.~A. Tainov,
  ``{Gravitational Waves Generated by Null Cosmic Strings},''
  \href{http://arxiv.org/abs/2311.01863}{{\tt arXiv:2311.01863 [gr-qc]}}.

\bibitem{Bidussi:2021nmp}
L.~Bidussi, J.~Hartong, E.~Have, J.~Musaeus, and S.~Prohazka, ``{Fractons,
  dipole symmetries and curved spacetime},''
  \href{http://dx.doi.org/10.21468/SciPostPhys.12.6.205}{{\em SciPost Phys.}
  {\bf 12} (2022) no.~6, 205}, \href{http://arxiv.org/abs/2111.03668}{{\tt
  arXiv:2111.03668 [hep-th]}}.

\bibitem{Figueroa-OFarrill:2023vbj}
J.~Figueroa-O'Farrill, A.~P\'erez, and S.~Prohazka, ``{Carroll/fracton
  particles and their correspondence},''
  \href{http://dx.doi.org/10.1007/JHEP06(2023)207}{{\em JHEP} {\bf 06} (2023)
  207}, \href{http://arxiv.org/abs/2305.06730}{{\tt arXiv:2305.06730
  [hep-th]}}.

\bibitem{Figueroa-OFarrill:2023qty}
J.~Figueroa-O'Farrill, A.~P\'erez, and S.~Prohazka, ``{Quantum Carroll/fracton
  particles},'' \href{http://dx.doi.org/10.1007/JHEP10(2023)041}{{\em JHEP}
  {\bf 10} (2023)  041}, \href{http://arxiv.org/abs/2307.05674}{{\tt
  arXiv:2307.05674 [hep-th]}}.

\bibitem{Perez:2023uwt}
A.~P\'erez, S.~Prohazka, and A.~Seraj, ``{Fracton infrared triangle},''
  \href{http://arxiv.org/abs/2310.16683}{{\tt arXiv:2310.16683 [hep-th]}}.

\bibitem{deBoer:2023fnj}
J.~de~Boer, J.~Hartong, N.~A. Obers, W.~Sybesma, and S.~Vandoren, ``{Carroll
  stories},'' \href{http://dx.doi.org/10.1007/JHEP09(2023)148}{{\em JHEP} {\bf
  09} (2023)  148}, \href{http://arxiv.org/abs/2307.06827}{{\tt
  arXiv:2307.06827 [hep-th]}}.

\bibitem{Ciambelli:2023tzb}
L.~Ciambelli and D.~Grumiller, ``{Carroll geodesics},''
  \href{http://arxiv.org/abs/2311.04112}{{\tt arXiv:2311.04112 [hep-th]}}.

\bibitem{Chandrasekaran:2021hxc}
V.~Chandrasekaran, E.~E. Flanagan, I.~Shehzad, and A.~J. Speranza,
  ``{Brown-York charges at null boundaries},''
  \href{http://dx.doi.org/10.1007/JHEP01(2022)029}{{\em JHEP} {\bf 01} (2022)
  029}, \href{http://arxiv.org/abs/2109.11567}{{\tt arXiv:2109.11567
  [hep-th]}}.

\bibitem{Armas:2023dcz}
J.~Armas and E.~Have, ``{Carrollian fluids and spontaneous breaking of boost
  symmetry},'' \href{http://arxiv.org/abs/2308.10594}{{\tt arXiv:2308.10594
  [hep-th]}}.

\bibitem{Bekaert:2015xua}
X.~Bekaert and K.~Morand, ``{Connections and dynamical trajectories in
  generalised Newton-Cartan gravity II. An ambient perspective},''
  \href{http://dx.doi.org/10.1063/1.5030328}{{\em J. Math. Phys.} {\bf 59}
  (2018) no.~7, 072503}, \href{http://arxiv.org/abs/1505.03739}{{\tt
  arXiv:1505.03739 [hep-th]}}.

\bibitem{Morand:2018tke}
K.~Morand, ``{Embedding Galilean and Carrollian geometries I. Gravitational
  waves},'' \href{http://dx.doi.org/10.1063/1.5130907}{{\em J. Math. Phys.}
  {\bf 61} (2020) no.~8, 082502}, \href{http://arxiv.org/abs/1811.12681}{{\tt
  arXiv:1811.12681 [hep-th]}}.

\bibitem{Figueroa-OFarrill:2018ilb}
J.~Figueroa-O'Farrill and S.~Prohazka, ``{Spatially isotropic homogeneous
  spacetimes},'' \href{http://dx.doi.org/10.1007/JHEP01(2019)229}{{\em JHEP}
  {\bf 01} (2019)  229}, \href{http://arxiv.org/abs/1809.01224}{{\tt
  arXiv:1809.01224 [hep-th]}}.

\bibitem{Figueroa-OFarrill:2020gpr}
J.~Figueroa-O'Farrill, ``{On the intrinsic torsion of spacetime structures},''
  \href{http://arxiv.org/abs/2009.01948}{{\tt arXiv:2009.01948 [hep-th]}}.

\bibitem{Levi-Civita:1917pgo}
T.~Levi-Civita, ``{Notion of Parallelism on a Generic Manifold and Consequent
  Geometrical Specification of the Riemannian Curvature},'' {\em Rend. Circ.
  Mat. Palermo} {\bf 42} (1917)  173--204,
  \href{http://arxiv.org/abs/2210.13239}{{\tt arXiv:2210.13239 [gr-qc]}}.

\bibitem{Ashtekar:1981hw}
A.~Ashtekar, ``{Radiative Degrees of Freedom of the Gravitational Field in
  Exact General Relativity},'' \href{http://dx.doi.org/10.1063/1.525169}{{\em
  J. Math. Phys.} {\bf 22} (1981)  2885--2895}.

\end{thebibliography}\endgroup

\end{document}